# Comprehensive Analysis on the Vulnerability and Efficiency of P2P Networks under Static Failures and Targeted Attacks

Farshad Safaei[1]and Hamidreza Sotoodeh[2]

[1] Faculty of ECE, Shahid Beheshti University G.C., Evin 1983963113, Tehran, IRAN
f_safaei@sbu.ac.ir
[2] Department of Computer Engineering, Qazvin Islamic Azad University, Qazvin, IRAN
hr.sotoodeh@qiau.ac.ir

## *ABSRACT*

*Peer-to-peer systems are the networks consisting of a group of nodes possible to be as wide as the Internet. These networks are required of evaluation mechanisms and distributed control and configurations, so each peer (node) will be able to communicate with other peers. P2P networks actually act as the specific transportation systems created to provide some services such as searching, large-scale storage, context sharing, and supervisioning. Changes in configuration, possibly the resultant effects of faults and failures or of the natural nodes behavior, are one of the most important features in P2P networks. Resilience to faults and failures, and also an appropriate dealing with threats and attacks, are the main requirements of today's most communication systems and networks. Thus, since P2P networks can be individually used as an infrastructure and an alternative for many other communication networks, they have to be more reliable, accessible, and resilient to the faults, failures and attacks compared to the client/server approach. In this work on progress, we present a detailed study on the behavior of various P2P networks toward faults and failures, and focus on fault-tolerance subject. We consider two different static failure scenarios: a) a random strategy in which nodes or edges of the network will be removed with an equal probability and without any knowledge of the network's infrastructure, b) a targeted strategy that uses some information about the nodes, and in which the nodes with the highest degree have the most priority to be attacked. By static faults, we mean a situation where the nodes or components encounter some faults before the network starts to work or through its operation, and will remain faulty to the end of the work session. Our goal is to introduce various measures to analyzing P2P networks evaluating their vulnerability rate. The presented criteria can be used for evaluating the reliability and vulnerability of P2P networks toward both random and targeted failures. There is no limit to the number and types of failures, the presented measures are able to be used for different types of failures and even a wide range of networks.*

## 1. INTRODUCTION

Networks are the backbone of modern ages, since they provide an infrastructure as the functioning basis of economy, societies and other requirements. The concept of network can be actively seen in almost all sciences. Analytical network models are changing to a point of interest to different fields of study; some of these models employ graphical representations to achieve their goals. Modern networks are different in features such as size, scale and fields of activity. For instance, transportation networks which transfer people in cities, have larger scale congestion compared to social networks such as World Wide Web, Facebook, MySpace, LinkedIn, and Twitter which are aimed to be a place for people to communicate and being entertained [1]. Network analysis and modeling bring a good understanding of organizational principles and mechanisms of different communication systems. The feature of learning gives a rise to the networks' modernity, so they can behave as a media providing communications and economic affairs.

In recent years, P2P networks have rapidly developed and extended and turned to a vital part of cyber culture and space. All current P2P applications are based on the graph theory properties which determine the resilience to node failures and the efficiency of routings, and are developed based on an application layer overlay. P2P network graphs range from centralized approaches such as star-like trees (e.g. Napster) to the connected $k$ -node complex graphs such as *Chord* [2] and *CAN* [3].

Cost, scalability, security, integrity and fault-tolerance are the factors to be considered in creating any felicitous communication network. Fault-tolerance is a critical factor in any communication network. Thus, with a rapid growth of computer networks and critical usages of these networks there is a need for networks robustness, especially under malfunctions and targeted attacks. Increasing hackers, router malfunctioning and other unpredicted failures, has made reliability a very hot research area [4]. In addition, studying the effects of breakdowns on real-world networks has been of a great interest in recent researches.

A large portion of studies in recent years is devoted to the topological construction of networks. Features like connectivity and shortest path length, can give good information of networks vulnerability. But on the other hand, properties such as the information flow passes the network, the resulted costs, and the users' behavior before and after a failure or attack, are of a greater importance since they can help considering the economical and functional aspects of the network. Reasons for this tendency to the aforementioned features and also to studying on the field of robustness can be summarized in two mail goals:

- Designing new networks considering whether they are subjected to only failures, or failures and attacks.

- Protecting the available networks, by locating the critical nodes and then doing actions in order to decrease their sensitivity.

Thus, to have a better understanding of a network, it is required to study and analyze its structure and topology and to focus on the dynamics of that network through the information flow passing the links. There are many previous work on the reliability evaluation of a network and making different network structures robust to the random failures and targeted attacks [5]. It is clearly obvious that robustness is one of the requirements of a network.

This paper is written to fulfill the aforementioned goals and aims to analyze the robustness and vulnerability of P2P networks. The peers in a P2P network may be removed from the network due to a random or a targeted failure. A targeted failure locates a node with the highest degree of connectivity and intends to attack the node. The absence of such a node, highly affects the connectivity of the network. For example, a peer which publishes illegal or uncopyrighted files can be a target in an attacked. We study the behavior of the networks under the both kinds of failures and try to have a quantitative evaluation of node removal and failures in a network. Thus, we can investigate the vulnerability and robustness of the networks toward the attacks causing a reduction in the performance and resources.

This paper is organized as follows. Section 2 describes the main terms and definitions used throughout the paper. In Section 3, a brief introduction of the robustness of P2P networks is proposed, and the fault models, different attack approaches, and robustness criteria are introduced and studied. Section 4 presents the simulation method and the assumptions and then analyzes the results. Finally, Section 5 concludes the paper and proposes suggestions for future work.

# 2. TERMS AND DEFINTIONS

P2P networks studied in this paper, generally can be presented by an undirected unweighted graph such as $G = (V, E)$, where $V$ refers to the set of nodes, and $E$ refers to the set of edges. Multiple-connection is not allowed in this graph, meaning that each edge can connect only one pair of nodes and not more, and any pair of nodes can at most be connected by one single edge.

## 2.1. P2P Network Model

P2P networks have been recently considered as an efficient and robust underlying for distributed systems and applications. The reliability of the services implemented by P2P technologies is highly dependent on the dependability of the underlying system. Thus, the way such system's behavior are analyzed, and also the consequent effects of every single node or component failure and the repairing mechanisms applied by users on the networks functionality is a main matter of discussion.

**Definition 1** [6]: *Any P2P structure can be modeled by a graph* $G = (V, E)$*, where* $V$ *is a finite set of nodes, and* $E$ *is a relation defined on* $V$*. The finite sequence* $\prec v_1, \ldots, v_k \succ$ *is a finite path from* $u$ *to* $v$ *if* $v_1 = u$, $v_k = v$, $\forall i : 1 \le i \le k-1$*, and* $(v_i, v_{i+1}) \in E$.

# 3. P2P NETWORKS ROBUSTNESS

Robustness of graphs and different kinds of networks has been specially considered in recent researches. Finding the conditions in which the occurred failures cause the network to be disconnected or significantly decrease the performance, is one of the classic problems in this field. Current studies deal with the homogenous node failures and can range from removals of one single node to a complete network disconnection. In this process the focus is on static failures, such that the simultaneous failures of the nodes in a congested network, occur with the independent probability $P_f$. However, researches [7] showed that P2P networks are considerably robust to the node failures and have the ability to tolerate half (50%) of the failures without any sensible degradation in functionality. Failures in links and nodes can cause the network to be disconnected or significantly slows down the performance.

## 3.1 Analysis of the Network Robustness

In this paper, we have proposed an approach for measuring P2P networks robustness to static random failures and targeted attacks, based on simulation. Before addressing to this approach, we should introduce different measures for performance evaluation of the networks, each of these measure evaluate the network from a different view. To better understand the impact of different attacks on the networks, two classes of networks are considered in our study. The evaluated P2P models in the paper are listed as CAN [3], Chord [2], Hypergrid [8], and PRU [9]. In addition, Erdös-Rényi [10] and (*p*, *g*) [11], two different real-world network topologies are included in our simulation. Each of these models and networks has detailed descriptions and algorithms that we refer the interested readers to the References section at the end of the paper in order to avoid repetition and also due to the space limitation. The networks are investigated through an abstract graph-based framework, and it is assumed that the graphs are undirected and the edges denote the relations between network nodes.

### 3.2 Measures for the Network Robustness

Different measures have been proposed for performance evaluation of networks, and each has a different role in choosing the proportional network. But as comparing all the measures requires a lot of time and using extended resources, it is not possible for us to compare all the measures in this paper. So we have only chosen the main measures and considered them to evaluate the networks robustness. The main measures are as follows:

- *The Disconnection probability:* the probability of the network to be disconnected under random failures or targeted attacks (see Definition 2).

- *The Cost factor* [12]: It is a quantitative description for the graph $G$ with $N$ nodes and $k$ edges and is given by

$$\text{Cost}(G) = \frac{2k}{N(N-1)} \tag{1}$$

As the network cost increases, the number of paths and consequently the connectedness of the network also increase.

- *The local efficiency* [12]: this measure is defined according to the average efficiency of local sub-graphs of the network, and it can be evaluated in two steps:

  I. Extract the node $i$ and all its first level neighbors

  II. Removal of the node $i$ and all the edges of related sub-graph

$$E_{loc}(G) = \frac{1}{N}\sum_{i \in V} E(G_i) \tag{2}$$

in which

$$E(G_i) = \frac{1}{k_i(k_i-1)} \sum_{l \neq m \in V} 1/d'_{lm} \tag{3}$$

In the above equation, $G_i$ is the sub-graph containing the neighbors of the node $i$, after removal of $i$, and $k_i(k_i-1)$ denotes the maximum number of edges possible for $G_i$. It should be noted that $d'_{lm}$ is the shortest path between the nodes $l$ and $m$ in $G_i$.

- *The global efficiency* [12]: the global efficiency of each network stands for the connectedness level of all the network nodes. This measure can be calculated by the equation below:

$$E_{glob}(G) = \frac{1}{N(N-1)} \sum_{\substack{i,j \in V \\ i \neq j}} \varepsilon_{ij} = \frac{1}{N(N-1)} \sum_{\substack{i,j \in V \\ i \neq j}} 1/d_{ij} \tag{4}$$

In the above equation, $\varepsilon_{ij}$ denotes the efficiency between the two nodes $i$ and $j$ ; this parameter is also assumed to be inversely related to $d_{ij}$ , that is the path length. Thus, if there is no connection between the two nodes $i$ and $j$ , then $d_{ij} = \infty$ and $\varepsilon_{ij} = 0$ . It is worth to mention that Equation (4) is defined for the conditions in which the network is disconnected due to the failure or the attack; although it is a valid equation for both connected and disconnected modes, and have a value in the range [0, 1] meaning that it does not require normalization.

- *The size of the giant component* [12]: this measure is defined as the maximum number of nodes existing in the largest cluster, after applying a failure to the network and the resulting disconnection.

## 3.3. The Rewiring Strategies

There is a defense strategy for every attack which neuters or reduces the impacts of that attack on the network. Rewiring [13] is one of those strategies which helps the network to balances the effects of attacks by creating new connections among the nodes. In rewiring, the affected node, i.e., the node which one of its neighbor nodes has been attacked, selects a node at random with the probability *P* to replace with its affected neighboring node. It is possible for a graph to be changed from a regular configuration (*P*=0) to a random graph (*P*=1), with a change in the value of mentioned probability [11]. There are two different approaches of rewiring [13]. We have proposed a novel third strategy for repairing the network after applied random failures and targeted attacks, and have evaluated its performance alongside the other two strategies in the mentioned network.

### 3.3.1. The Random Rewiring

In this approach the affected node randomly selects another node of the network as a replacement [13]. Due to the random style of selection, the network will be reconnected and the global efficiency will be increased using this approach.

### 3.3.2. The Greedy Rewiring

This approach is set to behave in a local way; the affected node selects a node with the highest degree among its second neighbors and connects to the selected node [13]. Since the structure of this approach is local, it is not able to repair the network disconnection. But, it can increase the local efficiency. Actually, the rewiring approach is a proportional defensive algorithm which allows the network to go on the defensive towards the failures and attacks.

### 3.3.3. The Betweenness Rewiring

The betweenness rewiring as well as greedy strategy acts with considerations to the network parameters. As mentioned in Section 3.2, the betweenness coefficient shows the importance degree of a node in connections with other network nodes. The higher this coefficient is for a node, the more paths pass through this node to connect other nodes of the network. Every damaged node in this approach looks for a node with a higher betweenness coefficient among its second neighbors. Such a node is actually a communication center between the local nodes and other network nodes. For example, as seen in Figure 1 when the node $a$ is removed from the network due to an attack, the affected node $b$ as the neighbor of $a$ tries to start a new connection with a node having the highest betweenness coefficient. Thus, node $b$ starts a new connection with node $c$ according to betweenness rewiring strategy.

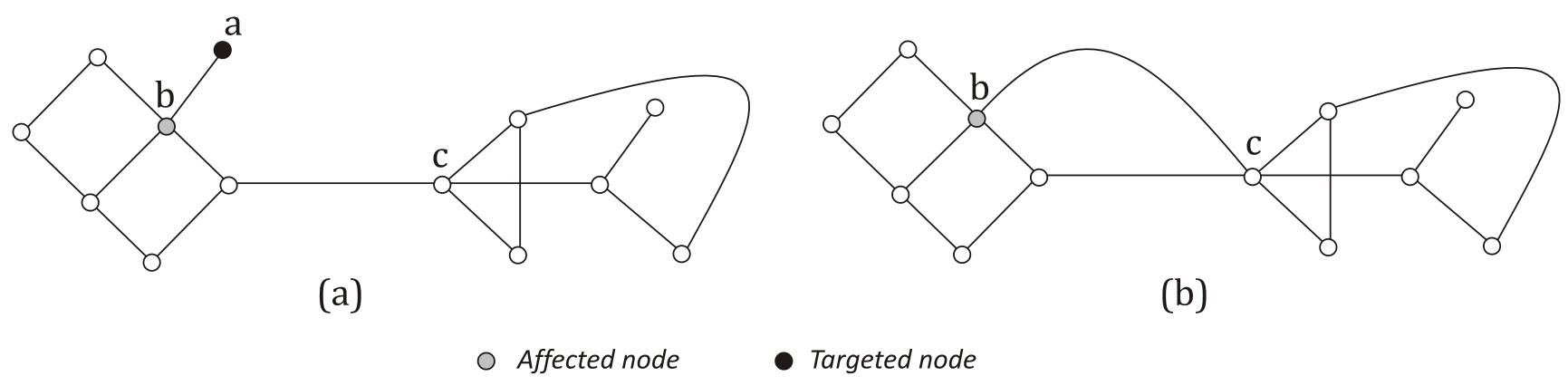

**Figure 1.** An instance network; (a) before applying betweenness rewiring; (b) after applying betweenness rewiring

## 3.4. The Fault Models

Most systems perceived as complex networks have actually a surprisingly degree of fault-tolerance which is the ability of the network and system allowing it to continue functioning despite the component failures [14]. This is a main outcome for evaluation and analysis of any network. An increase in the complexity of communication networks results into a rise in generated requests of the social media. Although the main components such as the routers and connection links can be failed on the network, the local failures can also help the network to lose the ability of carrying and delivering messages. The nodes may be removed from the network due to some faults and failures; this removals lead to a change in the global or local features of the network and its correct functioning as well.

Node failures and edge failures are two probable static fault models of the networks. In a node failure, all the connected edges will also fail and tagged as faulty. Faults can not only decrease the throughput and power of the network, but also may change the structure and topology of the network and result in a network disconnection.

**Definition 2** [7, 15]: *A network is called connected if there exists at least one path between every two healthy components; otherwise it will be called disconnected*.

Therefore, failures and attacks considering the selective removals of nodes or edges can decrease the performance of the network. This is generally a measure of performance degradation under failures, malfunctions and targeted attacks. The importance of studying the vulnerability of networks is in applying approaches to protect the networks towards failures and attacks. So, in order to protect the networks through isolation or quarantine one should be familiar with strategic approaches; and a good knowledge of robustness of vital networks is also required for creating robust networks.

In large-scale networks (e. g. P2P networks aiming at publishing uncopyrighted file copies) the whole network can be separated through targeted attacks towards the main nodes; this is another good reason for studying this field. It is interesting to know that the RIAA(Recording Industry Association of America) [13] is constantly searching for finding illegally published audios and videos, and tries to investigate and remove the publisher nodes in order to disintegrate the network.

### 3.4.1 The Attack Strategies

One of the important issues in network vulnerability studying is the method in which the nodes are under attacks. Each attack equals to a node removal in the network, which can damage the network structure and consequently the functioning of the network at different levels. Albert et al.

[16] have an investigation on both random and intentional removal of the nodes with higher degrees. Through their model they have assumed that the attacker tries to damage the network and remove the nodes. In random failure the nodes prone to fail or removed with the same probability. Intentional removals caused by targeted attacks have been organized to affect the nodes with the most connections; some number of clusters will be created in the network as a result of this kind of attack. Such attacks required previous information to explore the main nodes which itself takes times, and causes the time overload. Thus, in such attacks the well-informed attacker intentionally tends to cause damage in the network (as the vaccination function for removing bacteria, or an attack towards an illegal publisher node in a P2P file sharing network).

One of the attack strategies can be an arrangement of the nodes in a descending order of degrees and then start to remove the nodes on top of the list. Since this strategy uses the initial distribution of node degrees, it is called "ID-Removal". The nodes with the highest betweenness also play an important role in nodes connection. The betweenness measure is based on the concept of paths in the network, and the removals of main nodes can definitely affect the network. This measure for any node such $i$ states for a percentage of the shortest paths between other nodes which $i$ resides on [5]. In other words, we find the shortest paths between any pair of nodes and investigate to find out which percentage of the paths includes the node $i$. The symbol $C_B(i)$ denotes this parameter which is mathematically defined as [12]

$$C_B(i) = \sum_{\omega \neq \omega' \in V} \frac{\sigma_{\omega\omega'}(i)}{\sigma_{\omega\omega'}} \quad (5)$$

where, $\sigma_{\omega\omega'}$ is the shortest paths between $\omega$ and $\omega'$, passing through $i$. It should be noticed that the nodes placed on more number of shortest paths, have a higher betweenness.

The second strategy is called "IB-Removal" which uses the initial distribution of node degrees. Both ID- and IB-Removal strategies benefit from the initial information of the network. The more the nodes removed from the network, the more network topology alters and result in deviations in the degree distribution and betweenness from their initial values.

The third attack strategy is called "RD-Removal" which redistributes and recalculates the node degrees in each step. Finally, the fourth attack strategy which is based on recalculating the betweenness measure in each step is called "RB-removal". To sum it all up, it should be acknowledged that the first and third strategies are categorized as the local approaches, while the second and last strategies go in a global category. The local approaches try to rapidly remove the edges, while the global ones tend to destroy the shortest paths in the network.

In mentioned above malicious and targeted attacks it was assumed that the well-informed attacker have comprehensive information of the nodes and perform the attack based on the priorities. In addition, lack of information on the nodes is assumed in random failures. There is another attack strategy in which the attacker is only partly aware of the network components [17]. In other words, this strategy, which can be called as incomplete information based attacks, is different from the targeted attacks in a way that in a targeted attack providing incomplete information, the main nodes in a local area are the only components which will be attacked and consequently removed. This kind of attack will go between the two random and targeted attacks categories. The parameter $a$ indicates the attack information in this strategy; meaning that if there was no information of the network structure, it would be like $a = 0$ and the strategy is a random attack, while if a complete information was provided, it would be like $a = 1$ and the strategy is a targeted attack. Details on steps of attack with incomplete information strategy are reported in [17].

# 4. EXPERIMENTALS RESULTS

In order to evaluate P2P networks from the point of view of robustness, the assumptions of the evaluated networks should be presented at first. Below there are some assumptions considered to facilitate the evaluating presentation and the discussions on simulation. It should be noted that most of the models working on communication and P2P networks fault-tolerance field have been created considering the below assumptions [1, 6, 11-17].

- The network size: $N = 2000$ nodes.
- The mean node degree: $< k > = 18$.
- The networks types: CAN, Chord, Hypergrid, PRU, ER, and (*p, g*).
- Failure rate: $P_f$ = 0% ~ 80%.
- The attack strategies: Random, IB-Removal, ID-Removal, RB-Removal, RD-Removal, Incomplete Information.
- The rewiring mechanisms: Random, Greedy and Betweenness strategies with the probability *P*= 0.2, 0.4, 0.7.
- Only undirected networks are considered.
- The messages (the information flow) uniformly exchanged in the network.
- Random node failure means that the nodes follow an iid random process.
- The fault models are static; meaning that there is no fault occurrence during the calculation process. Of course, these fault models may disintegrate the network.
- The faulty components are unusable meaning that cannot transfer the information.
- The edges (links) in the network assumed to be reliable, while the nodes can be failed. This is an accepted assumption in many fault-tolerance literature [1, 6, 11-17]; the reason is the nodes are often more complex than the edges, so they have a higher probability of failure.
- Since in a P2P network the data belonging to the damaged node may be lost, there are several techniques among the nodes and peers of the network for data and information replication [1, 6, 11-17]. But as we focus on vulnerability of the network and not on restoring the lost information, we are not going to extend and study this aspect in our proposed plan.

## 4.1. The Simulation Method

Self-organizing, decentralization, and autonomy of P2P networks turn them to a proper candidate for distributed systems design. Since creating a real P2P network for performing the experiments requires to access and use many resources such as hundreds of users and computers, it is not able for us to test the system in reality before having it presented. Thus, simulation is the nearest method which can help us by evaluating the behavior of large-scale P2P networks. However the available simulators perform little flexibility on simulating the details of desired users, usages, and measures.

To evaluate the desired systems functioning, we have considered several P2P simulators from the points of views of functioning and network supporting. Different simulators which have been studied are as follows: Omnet [18], Peersim [19], NWB [20], and Pajek [21]. As the result of investigation, we found out that none of those simulators fulfills our desired metrics. We were also not able to reach some of the source files for insertion and edition purposes, however we needed to study the networks based on knowledge and the graph theory in different aspects such as disconnection. Thus, we decided to develop a software with the ability of simulating node type failures and removing the related connection links. This software can also provide us by evaluating the probability of disconnection and other desired measures. The simulation aims at measuring the probability of disconnection for some different percentages of faulty nodes in P2P networks and analyzing the resultant effects on the other network measures, and then comparing them with random and real-world networks. Each time running the simulator software it collects different statistics including: the probability of the network to be disconnected, the global and local efficiencies, the network cost, and the largest cluster size (giant component) after applying the failure.

The mentioned simulator is an object-oriented program implemented by C# language. Object-oriented design provide the ability to easily implement different network models with desired measures and random and targeted attack strategies, and evaluate the networks with new parameters. The fault model used in networks evaluation is a static model in which the nodes fail based on specific fault percentages at first, the result network will be evaluated then for the purpose of desired measures specifications. It is also possible to choose the rewiring mechanism and study the effects on repairing. The used simulation method is Monte-Carlo method; meaning that the statistics are collected for each run over 10 independent experiments. It is worth mentioning that in contrast with other methods there are no worries in this method for the simulator to reach a steady-state condition. The next subsection contains the simulation results and analysis.

#### 4.2.1 The Disconnection Probability

The importance of networks disconnection under random and malicious failures has been noticed earlier. The first analysis of this subsection is also devoted to this issue. Figure 2 indicates the probability of the networks to be disconnected under different attack strategies assuming the failure rate is set to 60%. At first glance it is clear that PRU network due to having a special graph structure which guarantees a low diameter for the network and increases the degrees of some nodes, is resilient only to the random attacks and has not been able to perform well under other types of attacks. The targeted attacks tend to attack the main and crucial nodes with the highest degrees or betweenness coefficients. So in a PRU network, the critical nodes with high degrees will be quickly removed and cause network disruptions.

A more accurate analysis of the figure makes us understand that all networks against IB-Removal attacks are weak and have high probabilities of disconnection. As mentioned before, the interconnection nodes which transfer the highest rate of messages are the targets in IB and RB attack types. Even structured networks such as CAN and Chord which are resilient to other attacks, will be damaged under these types of attacks. ($p$, $g$) network which is an improved model of scale-free networks, is more robust compared to the other networks and as it is clear in the figure it shows a good reaction to the most attack types and will be only disconnected under RD- and RB-Removal strategies. The reasons for this behavior can be justified such that the increase in parameter $p$ means to better the connectedness measure; and increasing both $p$ and $g$, means to increase the sensitivity of the network against targeted attacks. But, independently increasing of *p* or *g* is less effective.

We have chosen an easy solution to tune the parameters *p* and *g*, in a way that we have considered them as *p*=1 and *g*=0; so, having the (1, 0) we can achieve a network which is quite as connected as scale-free networks, and as fault-tolerant as ER model under targeted attacks. It should be noted that this network compared to the ER model, has a better performance under random attacks.

As expected, Hypergrid network is also well robust under ID- and RD-Removal attacks since it has limited node degrees, while its performance has been reduced against targeted attacks types 2 and 4. The behavior of ER random network is something to be mentioned in the figure. Although this network has a high probability of disconnection under the attacks types 2 and 4, it is quite robust to the other types of attacks. The studies of other measures reveal that it has also good global and local efficiencies compared to other networks.

An incomplete-information attack with selecting *a*=0.4 also couldn't bring a high disconnection to the networks. This attack has only caused a negligible disconnection, the same as ID-Removal attack, in Hypergrid network due to its special structure (limited node degrees).

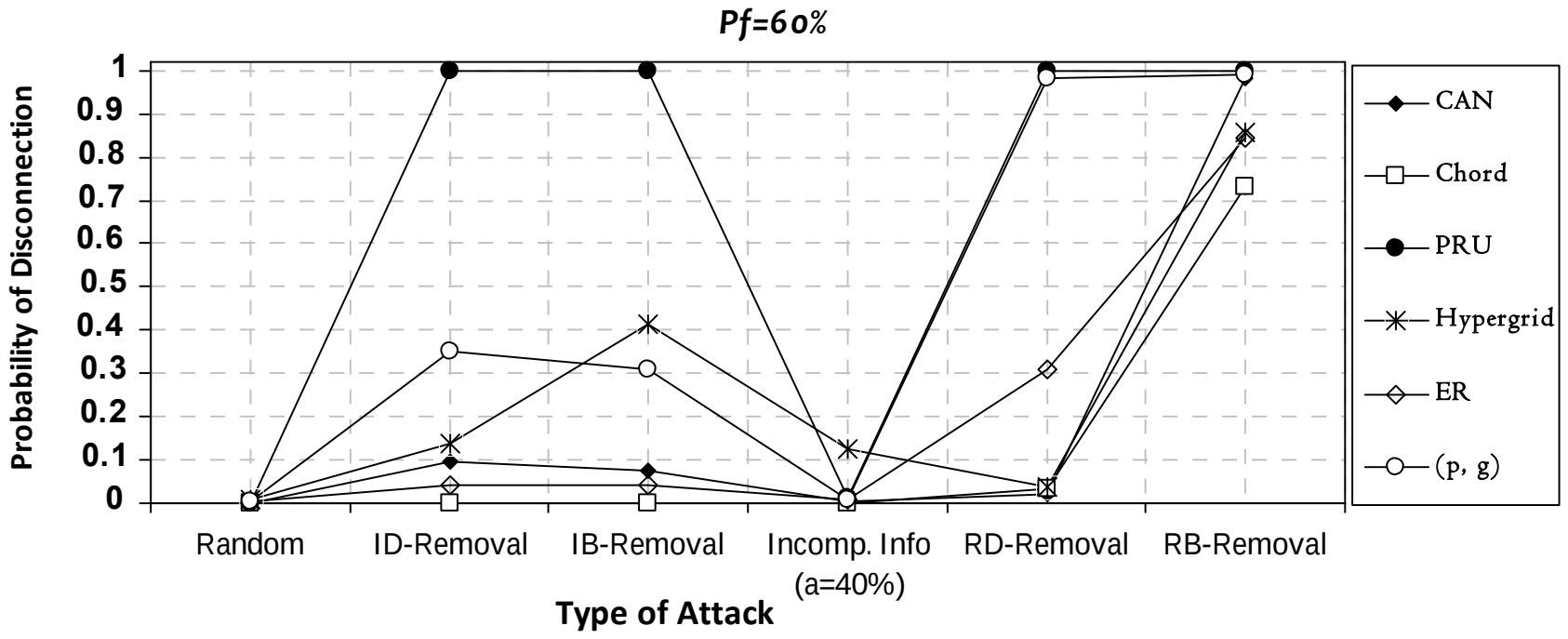

**Figure 2.** Comparison of the disconnection probabilities of different networks in terms of attack strategies in the presence of 60% failure rate ($P_f$=60%).

As mentioned before, all networks can be reconfigured by using random rewiring mechanism, but greedy and betweenness rewiring do not affect the probability of disconnection in the networks. Based on this facts we will continue with the studies on the impacts of rewiring on other measures.

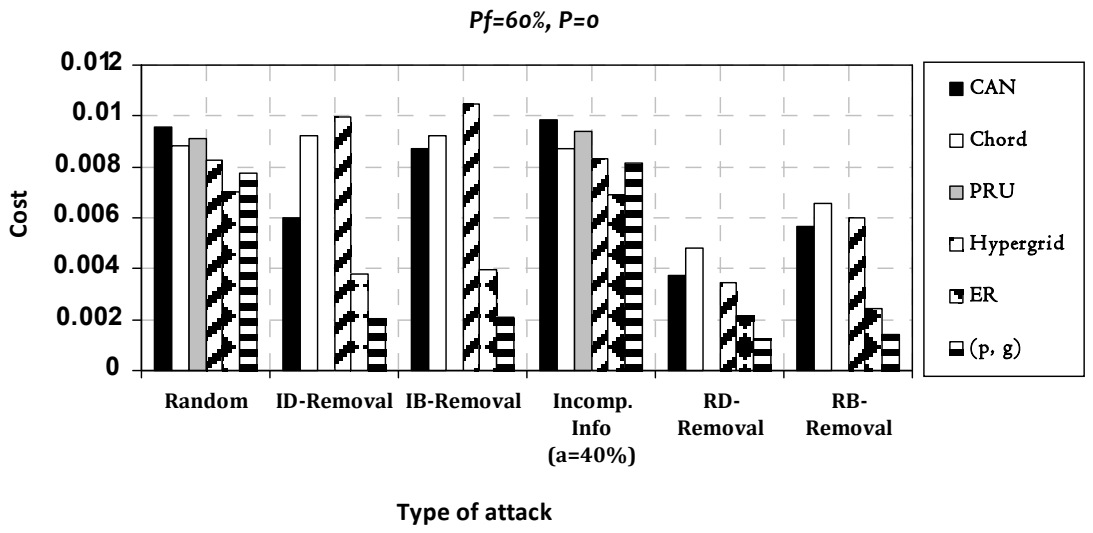

(a) Without the rewiring (*P*=0)

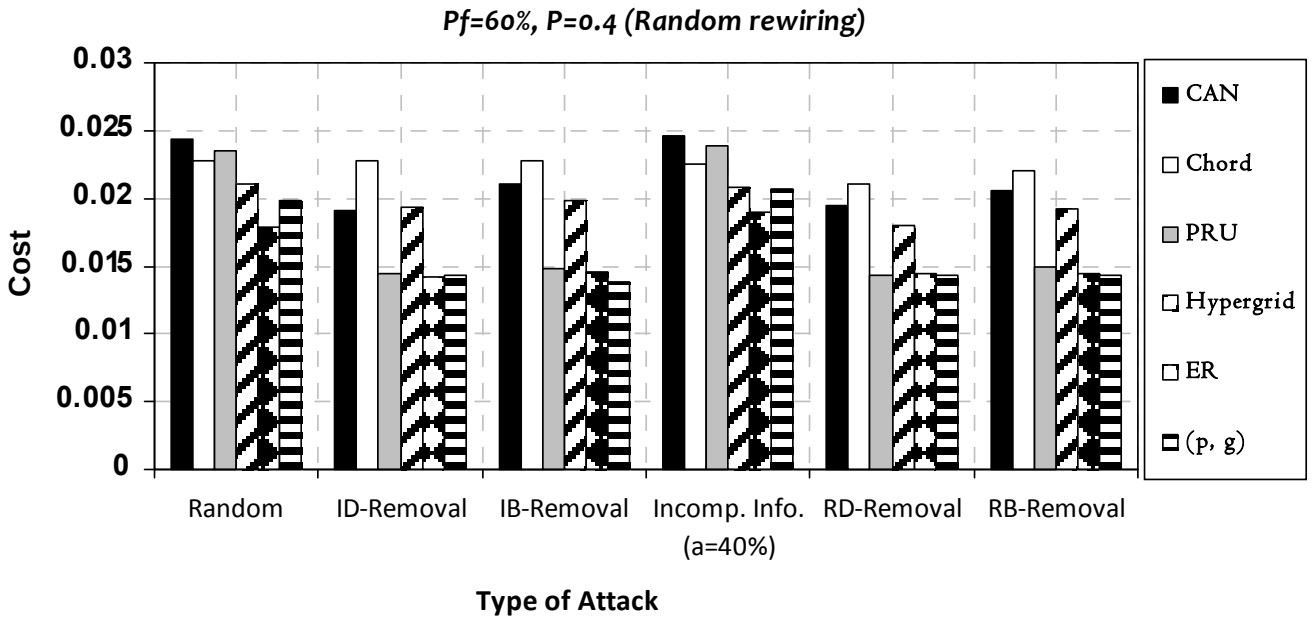

(b) With the random rewiring ($P$=0.4)

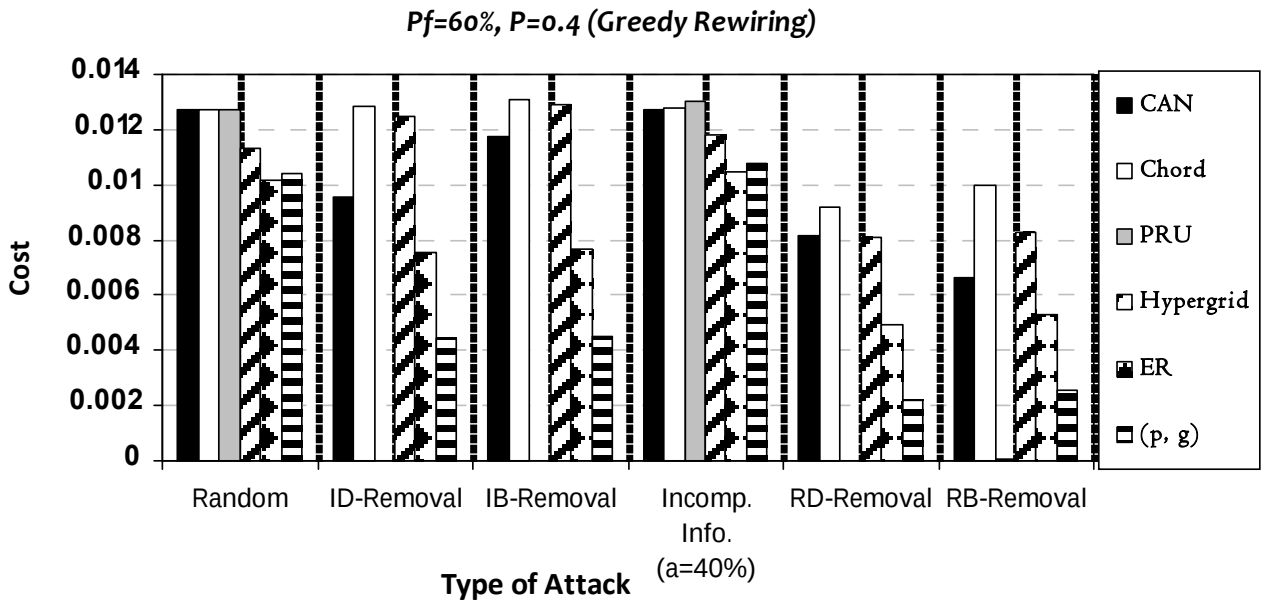

(c) With the greedy rewiring ($P$=0.4)

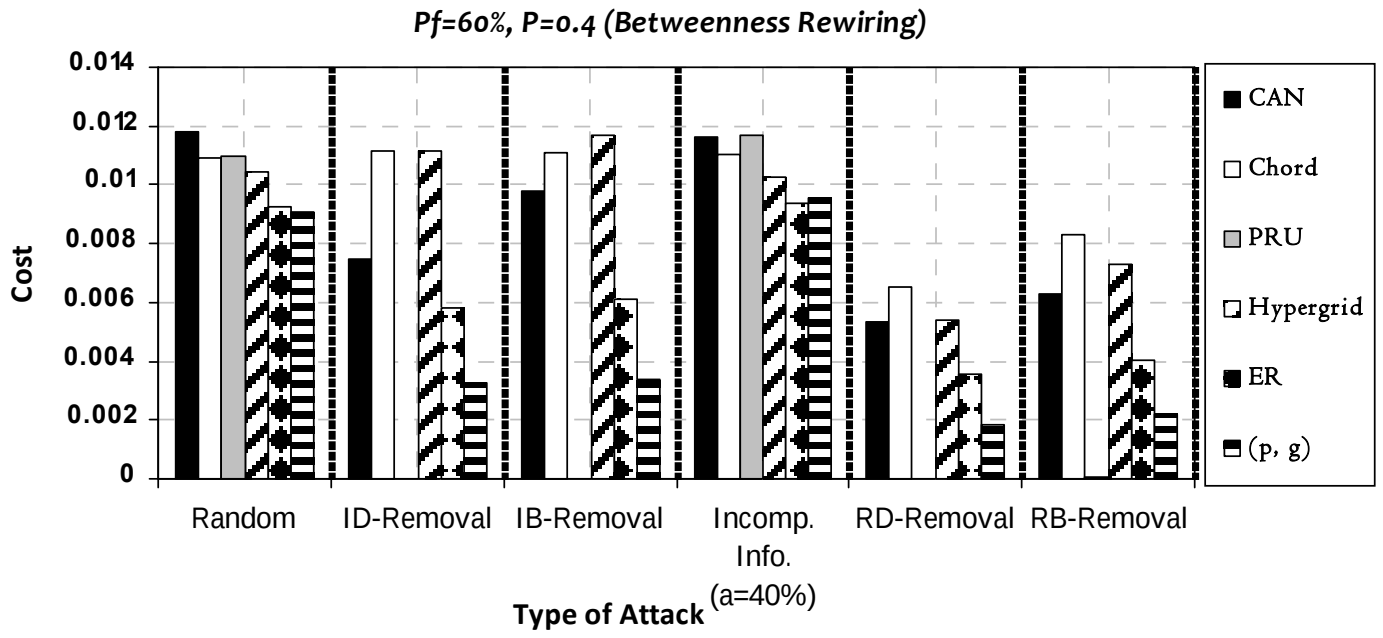

(d) With the betweenness rewiring ($P$=0.4)

**Figure 3.** Comparison of the cost factor of the networks in terms of the attack type using different rewiring strategies (the static failure rate ($P_f$) is set to 60% and the rewiring probability ($P$) is set to 40%).

### 4.2.2 The Cost Factor

The cost factor has been illustrated in Figure 3 using the random and greedy rewiring (with the probability $P$=0.4) and according to different types of attacks in the presence of 60% failure rate ($P_f$ =60%). The cost factor shows the level of connections in a network. The more the network cost is, the more probable the creation of shorter paths between nodes will be, and the global and local efficiencies will also increase as a result. As seen in Figure 3(a), the lowest cost goes to RD-Removal attack (which orders the nodes by their degrees at each step and then selects the node with the highest degree). The highest cost is also shown in the attack with incomplete information. The cost decreases through moving from random attacks to targeted attacks; it is due to the cost factor related equation which includes the square of network node count in the denominator. Node type attacks would definitely have greater impacts on this fraction.

Hypergrid network has the highest cost under most attacks, while PRU has the lowest cost. A network such PRU network which has high degree nodes, quickly loses its edges under targeted attacks and gets a significant decrease in the cost. While networks such as Hypergrid and Chord due to a balanced distribution of edges among the nodes, lose a less number of edges under different attacks and gets disconnected later.

The network cost after a greedy rewiring has been indicated in Figure 3(b). As expected, applying this mechanism does not cause the network a high cost; since it works locally and connects each node to its second neighbor having the highest degree. Thus, the isolated nodes or the nodes which would not be having a second neighbor after the failure has been applied are not able to create new links to the other nodes. After applying a greedy rewiring, PRU network in which the number of isolated nodes are high and is not actually affected by this mechanism would be having the lowest cost. The diagram indicates that the networks with a balanced and unified degree distribution would be mainly affected by this mechanism, while the random rewiring (Figure 3 (c)) has been significantly affected the cost in such networks. The reason has been laid in new connections between each node and another random node of the network. So that a network such PRU which is completely disconnected under targeted attacks, will be reconnected after applying a rewiring and get an increase in the cost.

The cost factor after applying the betweenness rewiring on different evaluated networks has been demonstrated in Figure 3(d). It shows that not only the betweenness rewiring does not have great impacts the cost factor, but also it has fewer effects compared to the greedy rewiring. The damaged node in Greedy strategy connects to the node having the highest degree, while nodes in Betweenness strategy start connections based on the betweenness coefficient. Thus, since the denominator in the cost factor fraction includes the square of node count, the greedy rewiring will has greater effects on the cost.

### 4.2.3 The Global Efficiency

Global efficiency, which shows the level of connections between networks nodes, is another important measure in a network. A higher global efficiency value shows a better connection among the network nodes and clusters. Figure 4 indicates the global efficiency measure according to $P_f$ =60% and three different rewiring strategies, random, greedy, and betweenness with the probability of $P$=0.2. Moving toward the targeted attacks through this evaluation also decreases the global efficiency. However the networks with a more balanced degree distribution have a higher global efficiency compared to the other networks; Chord, Hypergrid, and ER can be categorized as this type of networks. Such networks contain lots of minimum paths between the nodes which help them in continuing to work by quickly repairing their structure with replacing new paths while failures occur. CAN network is in the second place after the random ER network in terms of efficiency; which shows that this network divides to more clusters under failures. However in the following we will notice that CAN network has a significantly different local efficiency compared to the other networks.

The simulation results show that rewiring increases the global efficiency as well as the other measures. Each of these approaches improves the global efficiency in a different way. The greedy rewiring due to connecting the nodes to their second neighbors decreases the shortest path between the nodes; this can positively affect the global efficiency.

Using the random rewiring causes a network to be connected in a higher level. So some newly created paths appear in the network resulting in an increase in the global efficiency. That is a description of how random rewiring affects the global efficiency. In all the evaluated networks except Chord, using random rewiring increases the global efficiency in a larger scale compared to the greedy rewiring. The logarithmic structure of Chord which makes this network more resilient to failures is the reason for such a result. As seen, the probability of Chord network to be disconnected under attacks is much less compared to the other networks and it remains connected. So, greedy rewiring specifically decreases the length of shortest paths between the nodes and causes the global efficiency to be increased. This phenomenon can be seen in all the networks under random attacks. Moreover, it should be noted that all the networks remain connected under random attacks with a high probability. That makes it possible for the nodes to select a more number of neighbors for the purpose of replacing with the failed neighbor.

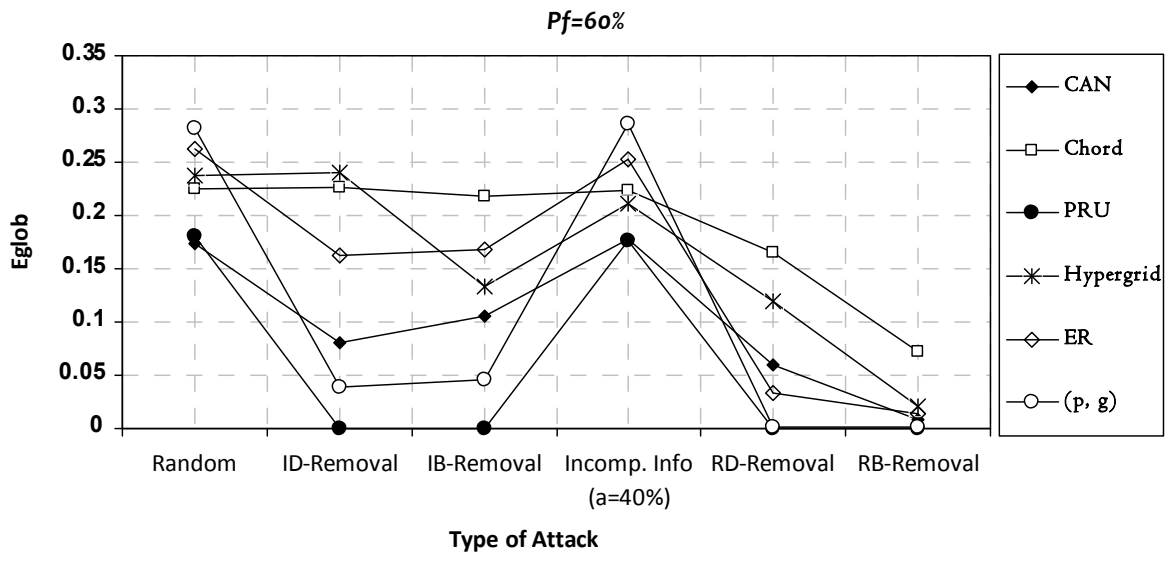

(a) Without the rewiring ($P$=0)

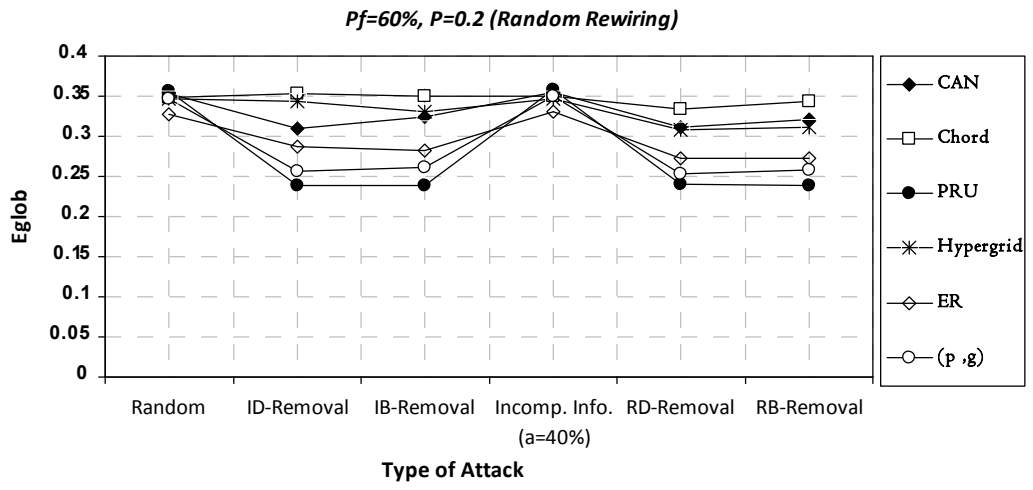

(b) With the random rewiring ($P$=0.2)

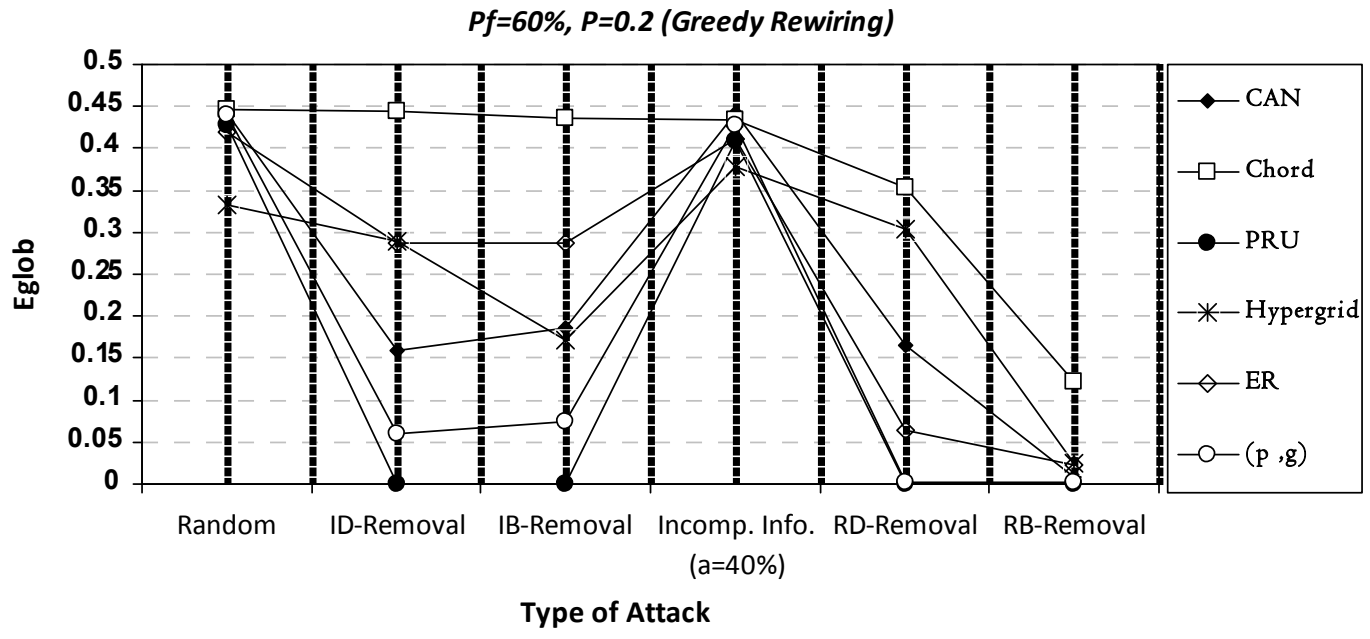

(c) With the greedy rewiring ($P$=0.2)

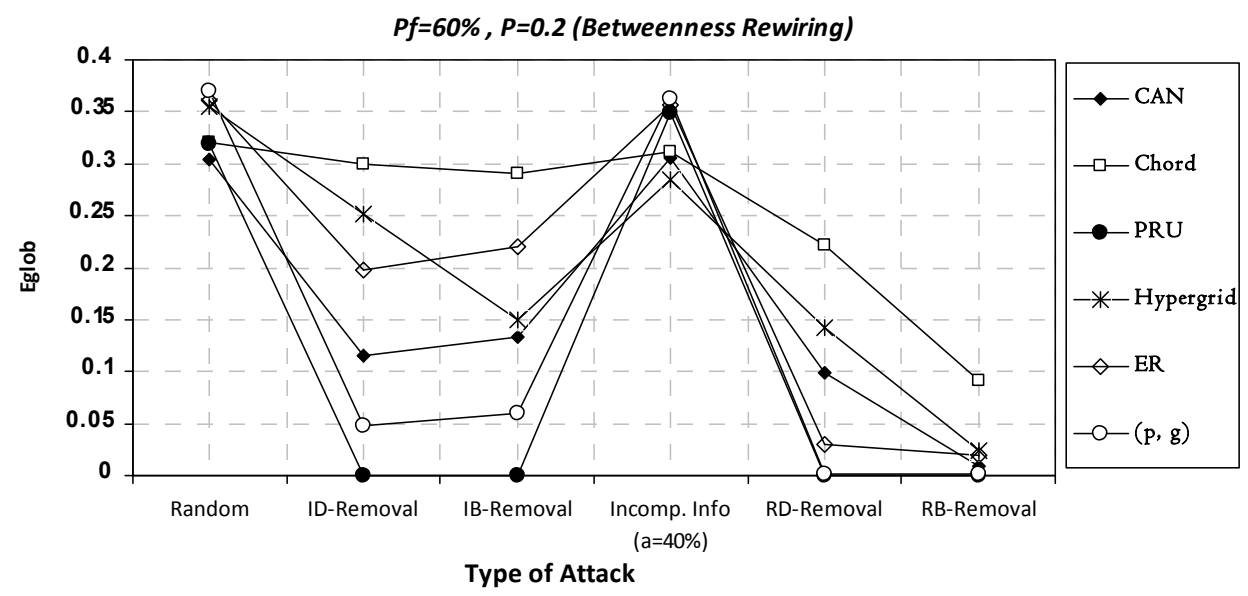

(d) With the betweenness rewiring ($P$=0.2)

**Figure 4.** Comparison of the global efficiency of the networks in terms of the attack type using different rewiring strategies (the static failure rate ($P_f$) is set to 60% and the rewiring probability ($P$) is set to 20%).

Thus, the probability of connecting to a node having a higher degree will be increased. It should be noted that the betweenness and greedy rewiring strategies contain the locality feature and have negligible impacts on the measures such as the global efficiency, the probability of disconnection, and the cost. Thus, considering Figure 4(d), it can be induced that this strategy has negligible impacts on the global efficiency.

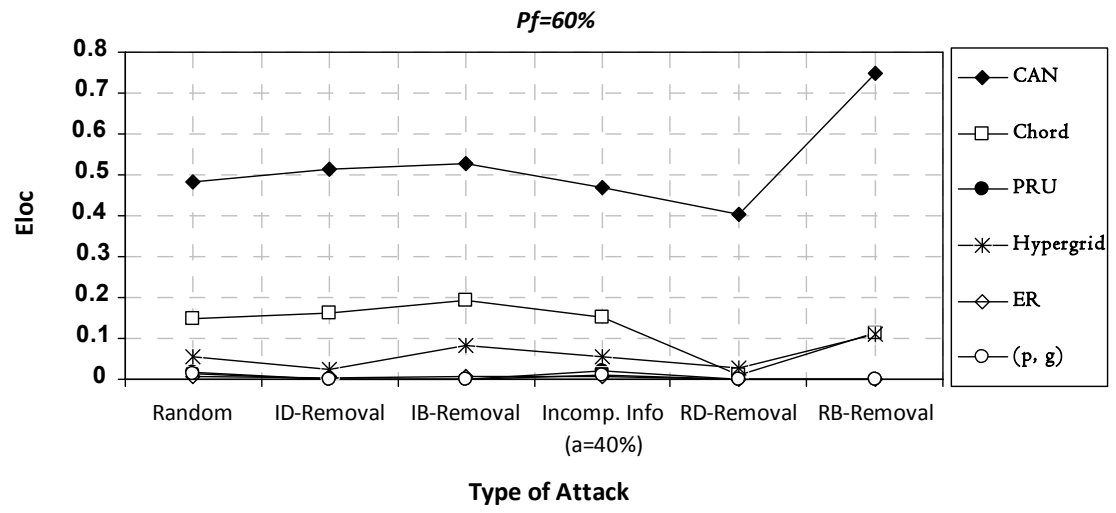

(a) Without the rewiring ($P$=0)

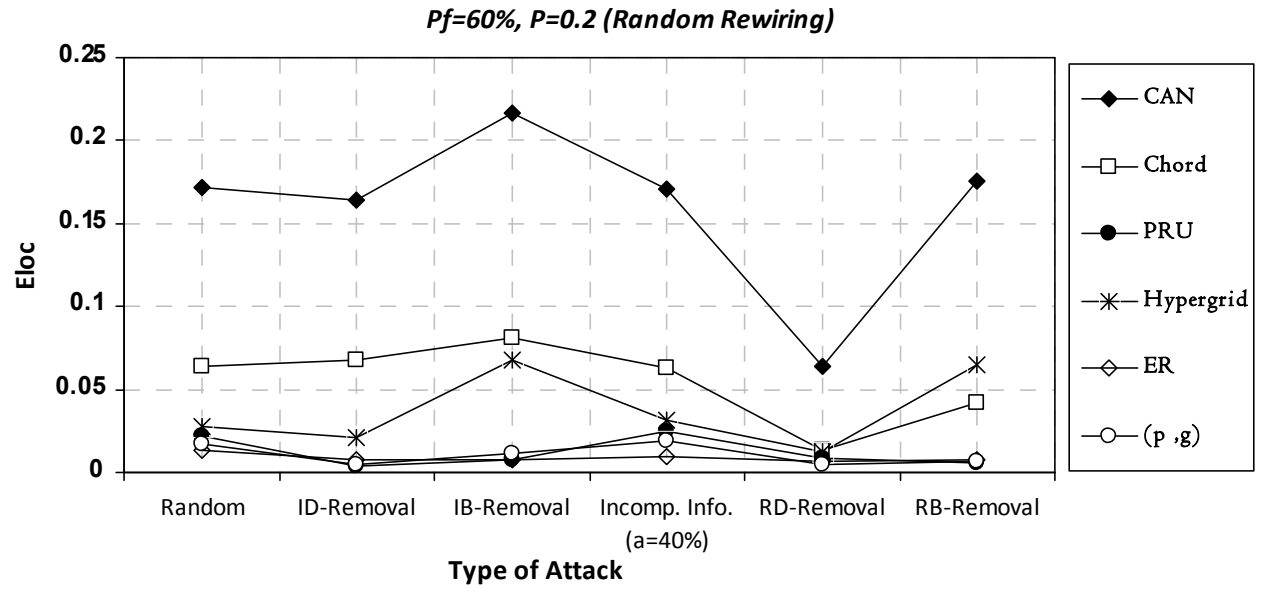

(b) With the random rewiring ($P$=0.2)

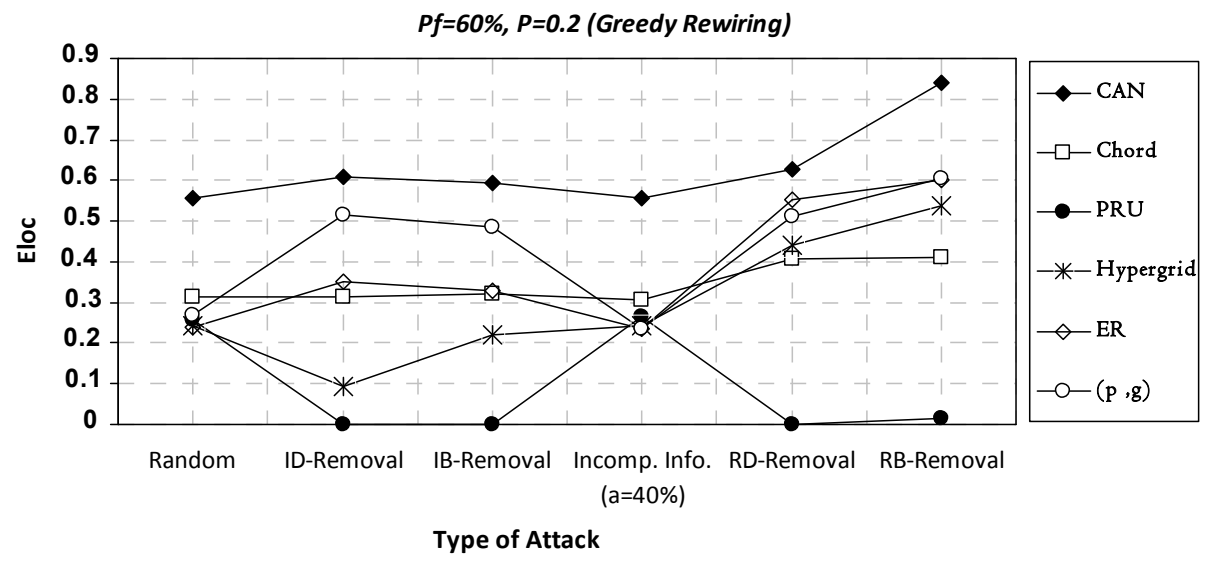

(c) With the greedy rewiring ($P$=0.2)

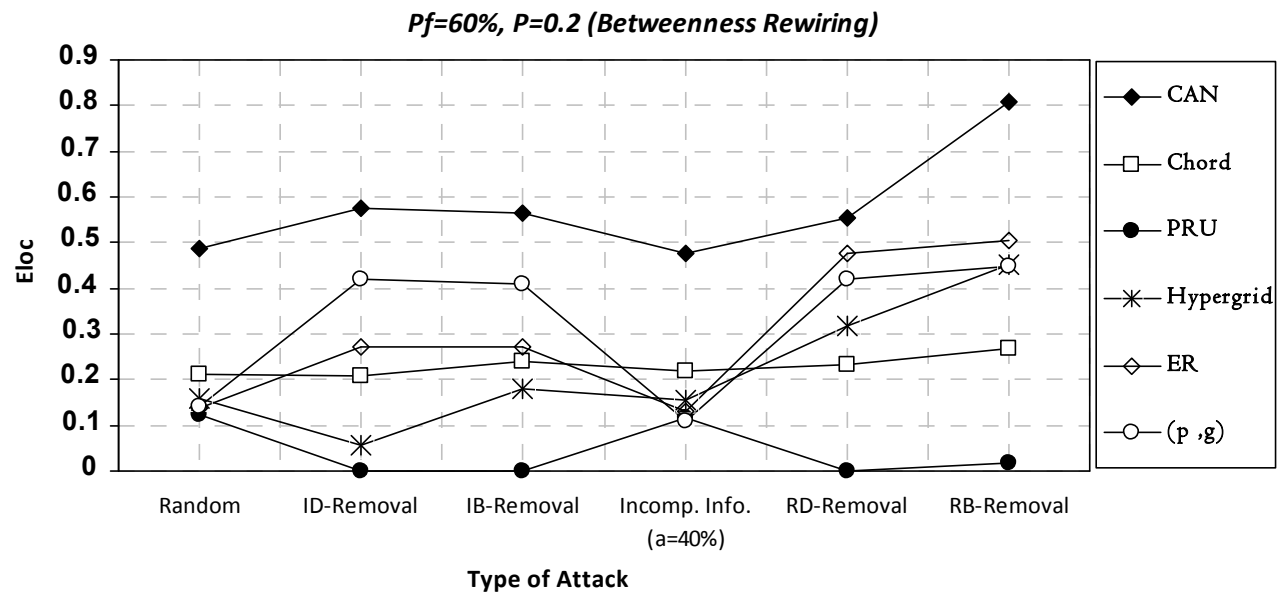

(d) With the betweenness rewiring ($P$=0.2)

**Figure 5.** Comparison of the local efficiency of the networks in terms of the attack type using different rewiring strategies (the static failure rate ($P_f$) is set to 60% and the rewiring probability ($P$) is set to 20%).

### 4.2.4. The Local Efficiency

In another experiment showing in Figure 5 we have studied the local efficiency for different types of networks in the presence of 60% failure rate with the impact of applied rewiring ($P$=0.2). As at the first glance it can be induced from Figure 5(a), CAN network has a higher local efficiency compared to the other networks; meaning that CAN has such a degree distribution and network structure that can show a good local efficiency even under applied random and targeted attacks. Studying the curves more accurately, we can understand that networks such as ER, PRU, and ($p$, $g$) have such lower local efficiencies which cannot be compared with other networks. In such networks each node compared to the all nodes has a lower degree; this issue results in a reduction in the local efficiency.

Another important point in studying the diagram is that the local efficiency shows the fault-tolerance feature of the network. The higher this measure is, the more fault-tolerant the network gets and it gets more resilient to failures and attacks. Thus, P2P networks compared to the real-world and random networks are more fault-tolerant against different types of attacks (random or targeted) and can better tolerate their users/nodes departures.

It is clearly obvious that using the greedy rewiring has positively affected the local efficiency. But something to be mainly noted is that the random rewiring has been effective in two different ways. A more accurate study of the diagram demonstrates that the random rewiring causes a negligible increase in the local efficiency of networks such as ER and ($p$, $g$), while has been significantly decreased this measure in some other networks such as CAN and Chord. It is due to the structural features of each network and the square of the maximum degree factor in the sub-network $G_i$ which is in the denominator of the related fraction (Equation (2)).

As seen in the figures, by applying the betweenness rewiring strategy the local efficiency has been significantly improved in all networks; it is because this strategy tries to select a node among its second neighbors having a higher communication factor with the cluster nodes. This issue makes the existing cluster paths length shorter which leads to increase the local efficiency quantity.

CAN and Chord are structured networks [2, 3] and use a special algorithm for connecting to the other nodes. For instance, a node in a CAN is not able to connect to any favorite nodes and can only connect to its neighbors in a *d*-dimensional space. So, the connections in such network are more local. This is while nodes in ER network can randomly select any favorite nodes in the whole network space to connect, so the nodes communication span is focused on the whole network. Considering that the local efficiency is inversely related to the maximum degree of the sub-network $G_i$, and also since in random rewiring the neighbors of each node are selected among the whole nodes, it can be said that when a new neighbor is added to the sub-network $G_i$, the maximum degree parameter in $G_i$ increased by a power of 2, while in a CAN or Chord network due to local connections there is not necessarily a path between the randomly selected node and the other neighbors of $i$ . As a result, the local efficiency decreased with an order of power 2. But in ER network, in which the connections span has been distributed through the network, the existence probability of a connection between the selected node and the neighbors of $i$ becomes higher and the local efficiency can be increased as a result. It should be noted that the sub-network $G_i$ contains all the neighbors of $i$ except $i$ itself.

### 4.2.4. The Size of the Giant Component

Failures in a network cause the network to be divided to smaller clusters with different number of nodes. If the number of nodes for each cluster is high enough, the network can still continue to work. The size of the giant component (the largest cluster size), $< S >$, is especially important among the clusters and shows the ability of the network for restarting its operation. In this section, we study the networks in terms of the number of nodes existing in the largest cluster after applying failures.

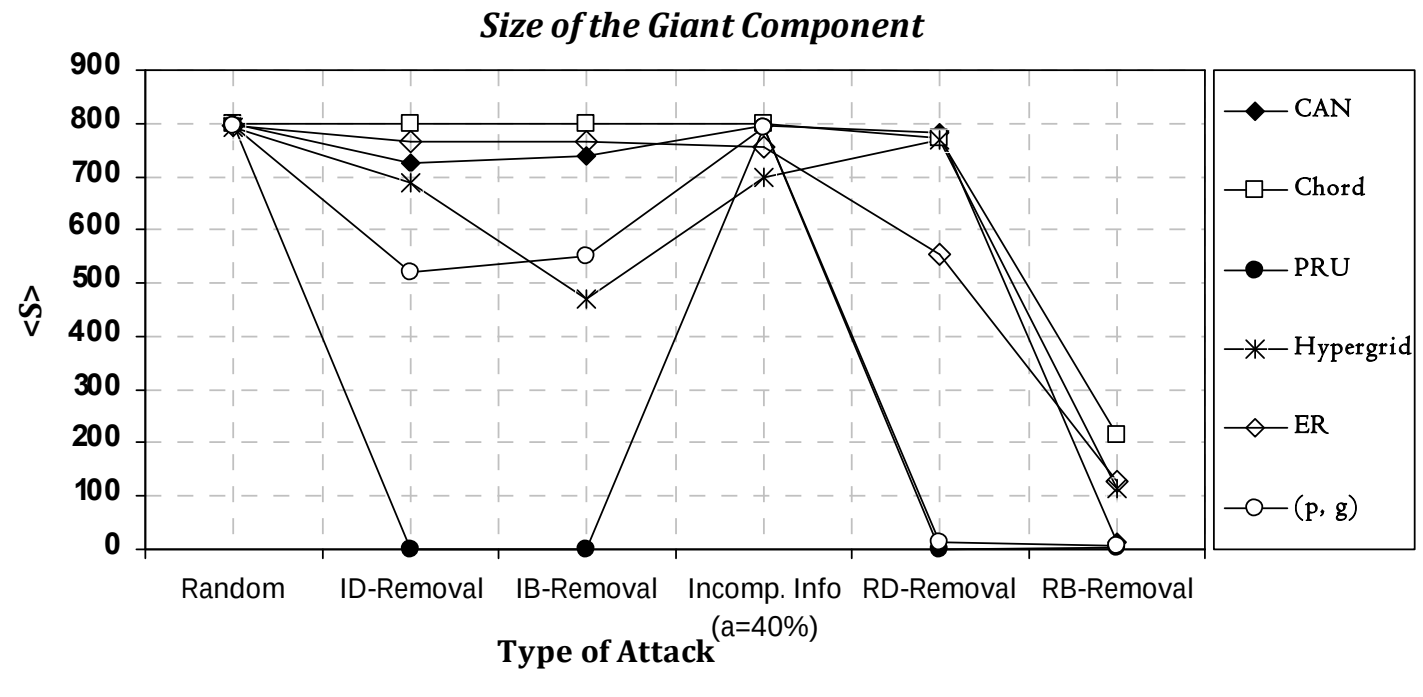

**Figure 6.** The size of the giant component with 60% failure rate ($P_f$ =60%).

Figure 6 compares the networks according to their largest clusters size and with the assumption of 60% failure rate ($P_f$=60%). The studies show that this measure is also sensitive to the targeted attacks, and it decreased while the attacks move towards being more targeted. It means that the network has been divided to more clusters. The diagram indicates that the largest cluster belongs to Chord network which has been resilient to the most attacks and still connected. The lowest size which is 1 belongs to PRU network showing that most probably there will be actually no connection between the nodes after applying targeted attacks. RB-Removal has been the most effective on the largest cluster among the attacks. In this attack the nodes are considered which have higher betweenness coefficients and actually play the role of bottlenecks. The loss of these nodes causes the network to lose its main connections and break down.

Before we mentioned that the greedy rewiring is not able to make the network connected due to the locality feature, and as a result it will not impact on the largest cluster size. On the other hand, the random rewiring affects the whole network and employing this mechanism would result in the network's connectedness. So, the largest cluster size would be equal to the number of network's healthy nodes.

# 5. CONCLUSIONS AND FUTURE WORKS

In this paper, by the help of introducing some criteria and conducting some simulations, we have studied the vulnerability of peer-to-peer (P2P) networks. We tried to analysis the impacts of faults and node failures in the static faults model and study their functionality in terms of the robustness in such networks. The term static robustness actually shows the ability of P2P network in resource assigning through the transition period between the node failures and repairing the whole system. The importance of this measure is of how quick each repairing mechanism does and finishes its work. The system which has a low static robustness needs a more rapid repairing technique. Although the static robustness is an important measure, it is not able to explain the network robustness in terms of attacker and malicious nodes and other damages. It can be considered as an issue to work on the future. In addition, we have proposed a novel rewiring mechanism named betweenness rewiring which we have evaluated its performance alongside both random and greedy rewiring mechanisms as defensive strategies to failures and attacks; we have also showed the impacts of this new mechanism on balancing the failure effects on the network. The simulation results have shown that using the random rewiring mechanism causes the network to be connected and improves the global parameters of the network, while it may have negative effects on the local parameters. On the other hand, using the greedy rewiring mechanism showed that there are not actually much effects on the global parameters; but it mostly affect the local parameters. The simulation results also showed that the local parameters in all networks would be significantly improved by employing the betweenness rewiring because the betweenness coefficient is the measure for creating connections in this mechanism. Finally, studying the evaluation results noted that the structured P2P networks can be resilient to a high rate of random failures and attacks and still remain connected. Important researches in line with the continuation of the present paper can be done on studying the vulnerability of the networks under dynamic faults. In such model the users (nodes) are able to leave the network according to their independent lifetime expiration. In order to evaluate the P2P networks which are under significant rate of these sudden changes (nodes arrival/departure), the robustness of such networks in terms of dynamic failures can be worth to study.